\documentclass[12pt]{article}
\usepackage{epsfig,graphicx}
\usepackage{fpcp03}

\def\beq{\begin{equation}}
\def\eeq#1{\label{#1}\end{equation}}
\def\eeqn{\end{equation}}

\def\beqa{\begin{eqnarray}}
\def\eeqa#1{\label{#1}\end{eqnarray}}
\def\eeqan{\end{eqnarray}}

\let\bar=\overbar

\def\etal{{\it et al.}}

\def\half{\frac{1}{2}}

\def\Dslash{\not{\hbox{\kern-4pt $D$}}}
\def\dslash{\not{\hbox{\kern-2pt $\del$}}}

\def\msb{{\bar{\ssstyle M \kern -1pt S}}}

\def\BB0bar{B^0 {\overline B}^0}
\def\BB0dbar{B_d^0 {\overline B}_d^0}
\def\BB0sbar{B_s^0 {\overline B}_s^0}

\RequirePackage{xspace}

\usepackage{relsize}
\def\babar{\mbox{\slshape B\kern-0.1em{\smaller A}\kern-0.1em
    B\kern-0.1em{\smaller A\kern-0.2em R}}}

\def\epem       {\ensuremath{e^+e^-}\xspace}

\def\qbar  {\ensuremath{\overline q}\xspace}

\def\dbar  {\ensuremath{\overline d}\xspace}

\def\sbar  {\ensuremath{\overline s}\xspace}

\def\cbar  {\ensuremath{\overline c}\xspace}

\def\b     {\ensuremath{b}\xspace}

\def\t     {\ensuremath{t}\xspace}

\def\piz   {\ensuremath{\pi^0}\xspace}

\def\pip   {\ensuremath{\pi^+}\xspace}
\def\pim   {\ensuremath{\pi^-}\xspace}

\def\Kbar  {\kern 0.2em\overline{\kern -0.2em K}{}\xspace}

\def\Kz    {\ensuremath{K^0}\xspace}
\def\Kzb   {\ensuremath{\Kbar^0}\xspace}
\def\KzKzb {\ensuremath{\Kz \kern -0.16em \Kzb}\xspace}
\def\Kp    {\ensuremath{K^+}\xspace}
\def\Km    {\ensuremath{K^-}\xspace}

\def\KpKm  {\ensuremath{\Kp \kern -0.16em \Km}\xspace}
\def\KS    {\ensuremath{K^0_{\scriptscriptstyle S}}\xspace} 
\def\KL    {\ensuremath{K^0_{\scriptscriptstyle L}}\xspace}

\newcommand{\etapr}{\ensuremath{\eta^{\prime}}\xspace}

\def\Dbar    {\kern 0.2em\overline{\kern -0.2em D}{}\xspace}

\def\Dz      {\ensuremath{D^0}\xspace}
\def\Dzb     {\ensuremath{\Dbar^0}\xspace}
\def\DzDzb   {\ensuremath{\Dz {\kern -0.16em \Dzb}}\xspace}
\def\Dp      {\ensuremath{D^+}\xspace}
\def\Dm      {\ensuremath{D^-}\xspace}

\def\DpDm    {\ensuremath{\Dp {\kern -0.16em \Dm}}\xspace}

\def\B       {\ensuremath{B}\xspace}
\def\Bbar    {\kern 0.18em\overline{\kern -0.18em B}{}\xspace}

\def\BB      {\ensuremath{B\Bbar}\xspace} 
\def\Bz      {\ensuremath{B^0}\xspace}
\def\Bzb     {\ensuremath{\Bbar^0}\xspace}
\def\BzBzb   {\ensuremath{\Bz {\kern -0.16em \Bzb}}\xspace}
\def\Bu      {\ensuremath{B^+}\xspace}
\def\Bub     {\ensuremath{B^-}\xspace}

\def\BpBm    {\ensuremath{\Bu {\kern -0.16em \Bub}}\xspace}

\mathchardef\Upsilon="7107
\def\Y#1S{\ensuremath{\Upsilon{(#1S)}}\xspace}

\mathchardef\Deltares="7101
\mathchardef\Xi="7104
\mathchardef\Lambda="7103
\mathchardef\Sigma="7106
\mathchardef\Omega="710A

\def\Deltabar{\kern 0.25em\overline{\kern -0.25em \Deltares}{}\xspace}
\def\Lbar{\kern 0.2em\overline{\kern -0.2em\Lambda\kern 0.05em}\kern-0.05em{}\xspace}
\def\Sigbar{\kern 0.2em\overline{\kern -0.2em \Sigma}{}\xspace}
\def\Xibar{\kern 0.2em\overline{\kern -0.2em \Xi}{}\xspace}
\def\Obar{\kern 0.2em\overline{\kern -0.2em \Omega}{}\xspace}
\def\Nbar{\kern 0.2em\overline{\kern -0.2em N}{}\xspace}
\def\Xb{\kern 0.2em\overline{\kern -0.2em X}{}\xspace}

\def\mes        {\mbox{$m_{\rm ES}$}\xspace}

\newcommand{\tev}{\ensuremath{\mathrm{\,Te\kern -0.1em V}}\xspace}
\newcommand{\gev}{\ensuremath{\mathrm{\,Ge\kern -0.1em V}}\xspace}
\newcommand{\mev}{\ensuremath{\mathrm{\,Me\kern -0.1em V}}\xspace}
\newcommand{\kev}{\ensuremath{\mathrm{\,ke\kern -0.1em V}}\xspace}
\newcommand{\ev}{\ensuremath{\mathrm{\,e\kern -0.1em V}}\xspace}
\newcommand{\gevc}{\ensuremath{{\mathrm{\,Ge\kern -0.1em V\!/}c}}\xspace}
\newcommand{\mevc}{\ensuremath{{\mathrm{\,Me\kern -0.1em V\!/}c}}\xspace}
\newcommand{\gevcc}{\ensuremath{{\mathrm{\,Ge\kern -0.1em V\!/}c^2}}\xspace}
\newcommand{\mevcc}{\ensuremath{{\mathrm{\,Me\kern -0.1em V\!/}c^2}}\xspace}

\def\mus  {\ensuremath{\rm \,\mus}\xspace}

\def\mus        {\ensuremath{\,\mu{\rm s}}\xspace}    

\def\calA{{\ensuremath{\cal A}}\xspace}

\def\ra                 {\ensuremath{\rightarrow}\xspace}

\def\pep2{PEP-II}

\def\gsim{{~\raise.15em\hbox{$>$}\kern-.85em
          \lower.35em\hbox{$\sim$}~}\xspace}
\def\lsim{{~\raise.15em\hbox{$<$}\kern-.85em
          \lower.35em\hbox{$\sim$}~}\xspace}

\def\CP                {\ensuremath{C\!P}\xspace}

\def\stwob{\ensuremath{\sin\! 2 \beta   }\xspace}

\def\deltaz{\ensuremath{{\rm \Delta}z}\xspace}
\def\deltat{\ensuremath{{\rm \Delta}t}\xspace}
\def\deltamd{\ensuremath{{\rm \Delta}m_d}\xspace}

\xspace

\newcommand{\jprlBase}       {Phys.\ Rev.\ Lett.\xspace}
\newcommand{\jprBase}        {Phys.\ Rev.\xspace}

\newcommand{\jprl}      [1]  {\jprlBase\ {\bf #1}}
\newcommand{\jprd}      [1]  {\jprBase\ D~{\bf #1}}

\def\jetset74   {\mbox{\tt Jetset \hspace{-0.5em}7.\hspace{-0.2em}4}\xspace}

\providecommand{\goto}{\rightarrow}

\newcommand{\skz}{\ensuremath{S_{\etapr\KS}}}
\newcommand{\ckz}{\ensuremath{C_{\etapr\KS}}}
\newcommand{\ilam}{\ensuremath{{\cal I}m\lambda_f}}
\newcommand{\alam}{\ensuremath{\left|\lambda_f\right|}}

\providecommand{\etapr}{\ensuremath{\eta^\prime}}

\newcommand{\rSetapKz}{\ensuremath{0.02\pm0.34 \pm 0.03}}
\newcommand{\rCetapKz}{\ensuremath{0.10\pm0.22 \pm 0.03}}

\newcommand{\DE}{\ensuremath{\Delta E}}

\providecommand{\KS}{\ensuremath{K_S^0}}
\newcommand{\kzs}{\KS}

\providecommand{\UfourS}{\ensuremath{\Upsilon(4S)}}

\providecommand{\half}{\ensuremath{{1\over2}}}

\newcommand{\bit}{\begin{itemize}}
\newcommand{\eit}{\end{itemize}}
\newcommand{\bea}{\begin{eqnarray*}}
\newcommand{\eea}{\end{eqnarray*}}
\def\micron{\hbox{$\mu{\rm m}$}}
\newcommand{\plotdir}{plots}

\begin{document}

\newcommand{\BABARPubYear}    {03}
\newcommand{\BABARPubNumber}  {00}
\newcommand{\BABARProcNumber} {064}
\newcommand{\SLACPubNumber} {10153}
\newcommand{\LANLNumber} {0309001}

\begin{flushright}
\babar-PROC-\BABARPubYear/\BABARProcNumber \\
August 2003 \\
\end{flushright}



\Title{\boldmath Measurements of $\sin{2\beta}$ in $B$ decays}
\bigskip


%
\label{FordStart}

%
\author{ William~T.~Ford\index{Ford, W. T.} }

%
\address{Physics Department\\
University of Colorado \\
Boulder, Colorado 80309-0390, USA \\
}

\makeauthor\abstracts{
I describe the experimental determination of the CKM angle $\beta$ from
measurements of the decay time separation \deltat\ between \Bz\ mesons
produced as \BB\ pairs in \epem\ annihilation.  The current results from
leading and non-leading decay modes are presented and compared.
}

\section{Introduction}\label{sec:intro}
\newcommand{\svBz}{\ensuremath{|\Bz\rangle}}
\newcommand{\svBzb}{\ensuremath{|\Bzb\rangle}}
\newcommand{\svBzL}{\ensuremath{|B^{0\!}_L\rangle}}
\newcommand{\svBzH}{\ensuremath{|B^{0\!}_H\rangle}}
\newcommand{\svBzt}{\ensuremath{|B^{0\!}_{\rm phys}(t)\rangle}}

\newcommand{\lambdackm}{{\ensuremath{\lambda}}}
\newcommand{\lambdacp}{{\ensuremath{\lambda_f}}}

A primary objective of particle physics since the 1964 discovery of
\CP\ non-conservation in weak decays has been the understanding of
the effect as a natural feature of the structure of the weak
interaction.  A clue was offered in 1973 by Kobayashi and Maskawa
in anticipation of the discovery of a third fermion
family.  With three families the unitary transformation to the weak
isospin basis of the left-handed fermions has four parameters,
including one complex phase that breaks the \CP\ symmetry of the
flavor-changing transitions.  In the Wolfenstein parameterization the
CKM matrix $V$ is
\beq
  V = \pmatrix{
  V_{ud} = 1-\half\lambdackm^2 & V_{us} = \lambdackm & {V_{ub} = A\lambdackm^3(\rho-i\eta)}\cr
  V_{cd} = -\lambdackm     & V_{cs} = 1-\half\lambdackm^2 & {V_{cb} = A\lambdackm^2}\cr
  {V_{td} = A\lambdackm^3(1-\rho-i\eta)} & {V_{ts} = -A\lambdackm^2} & V_{tb} = 1 \cr},
\eeq{eq:ckm}
where $\lambdackm \simeq \sin\theta_c\simeq 0.22$ and $A \sim 1$.  The
unitarity equations, such as 
$V_{ud}V_{ub}^*+V_{cd}V_{cb}^*+V_{td}V_{tb}^*=0$, lead to ``unitarity
triangles'' in which the angle of interest here is 
\beq
\beta = \phi_1 = \arg{(-V_{cd} V^*_{cb}/V_{td} V^*_{tb})}.
\eeq{eq:betadef}

\newcommand{\Abar}{\kern 0.18em\overline{\kern -0.18em A}{}\xspace}

\section{\boldmath Time evolution of the decay and \CP\ %
violation}\label{sec:dt} 

For a neutral \B\ meson that is in the state \Bz\ at $t=0$ we write
the time-dependent amplitude for its decay to final state $f$ as
\beq
     \langle f | H \svBzt = e^{-imt}e^{-\Gamma t/2}\left[{A_f}\cos{\half\Delta m\xspace t} +
     i{\frac{q}{p}}{\Abar_f}\sin{\half\Delta m\xspace t}\right],
\eeq{eq:amplitudeB2f}
where $A_f \equiv \langle f | H | \Bz\rangle$ is the amplitude for
flavor-definite \Bz\ decay to the final state $f$, $\Abar_f$ is the
corresponding amplitude for \Bzb, and $p,q$ give the weak eigenstates
$B^{0\!}_{L,H}$ in the $(\Bz,\Bzb)$ basis:  $|B^{0\!}_{L,H}\rangle =
p|\Bz\rangle\pm q|\Bzb\rangle$.
The average and difference between $H$ and $L$ masses are $m$ and
$\Delta m$, respectively, and we've taken the approximation
$\Gamma_H=\Gamma_L=\Gamma$ for the decay rates. 
Particle-antiparticle mixing is responsible for the non-exponential
behavior.  When $f$ is accessible to both \Bz\ and \Bzb\ the violation
of \CP\ symmetry appears through the interference 
between mixing ($q/p$) and decay ($\Abar_f/A_f$), even
if \CP\ is conserved in both ($|q/p|=|\Abar_f|/|A_f|= 1$).
In \UfourS\ decay a \BzBzb\ pair is created in a $C=-1$ eigenstate, and
the two mesons oscillate coherently between \Bz\ and \Bzb\ until one decays
(Einstein-Podolsky-Rosen effect).  

In the experiments there is no
marker of $t=0$; rather one must consider the time separation \deltat\
between the decay of one \B\ to a flavor eigenstate (``{tag}''), and 
the decay of the other \B\ to the
\CP\ eigenstate $f$.  The resulting decay rate in terms of \deltat\ is
\beq
     \frac{d\Gamma^f_\pm(\deltat)}{d\deltat} \propto 
       e^{-|\deltat|/\tau}\left(1\pm{{\cal I}m\lambdacp}\sin{\Delta
     m\xspace\deltat}\right),
\eeq{eq:idealGamDt}
where $\lambdacp\equiv\frac{q}{p}\frac{\Abar_f}{A_f}$, we've
assumed $\alam=1$, and the $+(-)$ sign labels a \Bz(\Bzb) tag.  The
ability to measure \deltat\ depends on the motion of 
the \UfourS\ in the rest frame of the experiment, leading to direct
measurement of $\Delta z \simeq \beta\gamma c \Delta t$ ($z$ being the
boost axis).  The boost magnitudes for the two asymmetric \B\ factories
are $\beta\gamma=0.56$ for PEP-II, and $\beta\gamma=0.425$ for KEKB.

In the analysis one reconstructs the decay of one \B\ meson in the
final state $f$, assumed here to be a \CP-eigenstate.  Coming from
$\Upsilon(4S)\goto \BzBzb$ (or \BpBm), the \B\ meson is
nearly at rest $(p^*_B\simeq 325\ \mevc)$.  This leads to a strong
correlation between the reconstructed mass and the missing mass of the
partner \B.  The usual choice of an independent pair of kinematic
variables is
\beq
  \DE = E^*_B - E^*_{\mathrm{beam}}, \qquad
  \mes = \sqrt{E_{\hbox{beam}}^{*2} - \left|{\bf p}^*_B\right|^2}.
\eeq{eq:demb}
Here the asterisk denotes \UfourS\ frame, and the subscript $B$
denotes the reconstructed \B.  
One looks for central values of these
variables peaked at $\DE=0$ and $\mes=m_B$.  Typical resolutions are
$\sigma(\mes)\simeq 3\ \mevcc$ and $\sigma(\DE)\simeq15-50\ \mev$

The other (tag) $B$ is not fully reconstructed; we need to know 
only its decay point and whether it's \Bz\ or \Bzb.  The particles
left over from reconstruction of the $\B\ra f$ candidate are therefore
examined to form the recoil \B\ vertex and to deduce its flavor.  
The resolution on \deltaz, of which the largest contribution comes
from tag side, is $\simeq180\ \micron$,
or $\sim1.25$ ps, and is similar for \babar\ and Belle.

Flavor tagging signatures include the sign of charge in ($\Bz\goto
\ell^+,\ \Bzb\goto \ell^-$), ($\Bz\goto K^+,\ \Bzb\goto K^-$), leading
charged track, etc.  The efficiency $\epsilon$ and mistag rate $w$ for
each algorithm is measured with reconstructed flavor eigenstate decays.
The effective efficiency is given by $Q = \epsilon(1-2w)^2$.  The
average values of this figure of merit are $(28.1\pm0.7)\%$ reported
by \babar, and $(28.8\pm0.6)\%$ by Belle.

\begin{figure}[htbp]
\begin{center}
 \scalebox{.7}{
  (a) \includegraphics[width=.5\linewidth,bb=132 650 335 736]%
   {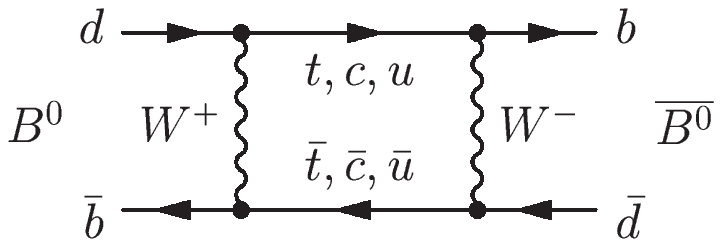}\qquad
  (b) \includegraphics[width=.39\linewidth,bb=132 650 294 732]%
   {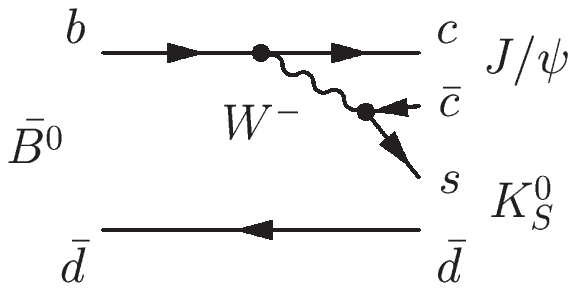}
 }
\end{center}
\vspace{-7mm}
 \caption{\label{fig:diagMix} (a) $\B\Bbar$ mixing diagram; (b)
leading diagram for $\B\ra {\rm charmonium}\ \kzs$ decays.}
\end{figure}
After convolution with the resolution function ${\cal R}$ and provision
for imperfect tagging the formula for the decay rate becomes 
\beq
     \frac{1}{\Gamma^f}\frac{d\Gamma^f_\pm(\deltat)}{d\deltat} =
       \frac{e^{-|\deltat|/\tau}}{4\tau}\left[1\pm(1-2w)
       {\cal I}m\lambdacp\sin{\Delta m\xspace \deltat}\right]
       \otimes{\cal R}.
\eeq{eq:realGamDt}
Thus we are looking for an asymmetry between \Bz- and \Bzb-tagged events
having a sinusoidal \deltat\ dependence with known angular frequency $\Delta m$
and amplitude given by ${\cal I}m\lambdacp$ multiplied by the known dilution
factor $1-2w$.  The physics is in ${\cal I}m\lambdacp$; it contains the
factor $q/p$, common to 
all decay modes, that can be calculated from the diagram in Fig.\
\ref{fig:diagMix}a.  With the heavy \b\ quark we have confidence in the
short-distance calculation at parton level.  The virtual \t\ quark
dominates in the loop, since  
its large mass is responsible for violating the GIM
mechanism that otherwise suppresses the mixing.  The result is
\beq
 \frac{q}{p} = \frac{V_{tb}^*V_{td}}{V_{tb}V_{td}^*},
\eeq{eq:qoverp}
which in the Wolfenstein phase convention is $e^{-2i\beta}$.

\section{\boldmath \stwob from charmonium $K^{0(*)}$ modes}\label{sec:chmmK}

The charmonium $K^{0(*)}$ decays are well described in terms of the
color-suppressed, CKM-favored tree diagram of Fig.\ \ref{fig:diagMix}b.
The ratio of amplitudes entering into \lambdacp\ is given by
\beq
      {\frac{\Abar_f}{A_f}} =
       \eta_f\left(\frac{V_{cb}V^*_{cs}}{V^*_{cb}V_{cs}}\right)
             \left(\frac{p}{q}\right)_K =
       \eta_f\left(\frac{V_{cb}V^*_{cs}}{V^*_{cb}V_{cs}}\right)
             \left(\frac{V^*_{cd}V_{cs}}{V_{cd}V^*_{cs}}\right)
     = {\eta_f},
\eeq{amplRatioChmmKs}
where $\eta_f=+1\ (-1)$ for a $CP$ even (odd) final state $f$, and the
last step assumes the Wolfenstein phase convention. 
Combining with Eq.\ \ref{eq:qoverp}\ we find, independent of phase
convention, 
\beq
   \lambdacp = \eta_f e^{-2i\beta},\qquad 
     {\cal I}m\lambdacp= -\eta_f\sin{2\beta}.
\eeq{eq:lambdaChmmKs}
Thus the amplitude of the sine term in Eq.\ \ref{eq:realGamDt}\ gives
directly \stwob.

\begin{figure}[htbp]
\scalebox{1}[.9]{%
 \includegraphics[width=.49\linewidth]{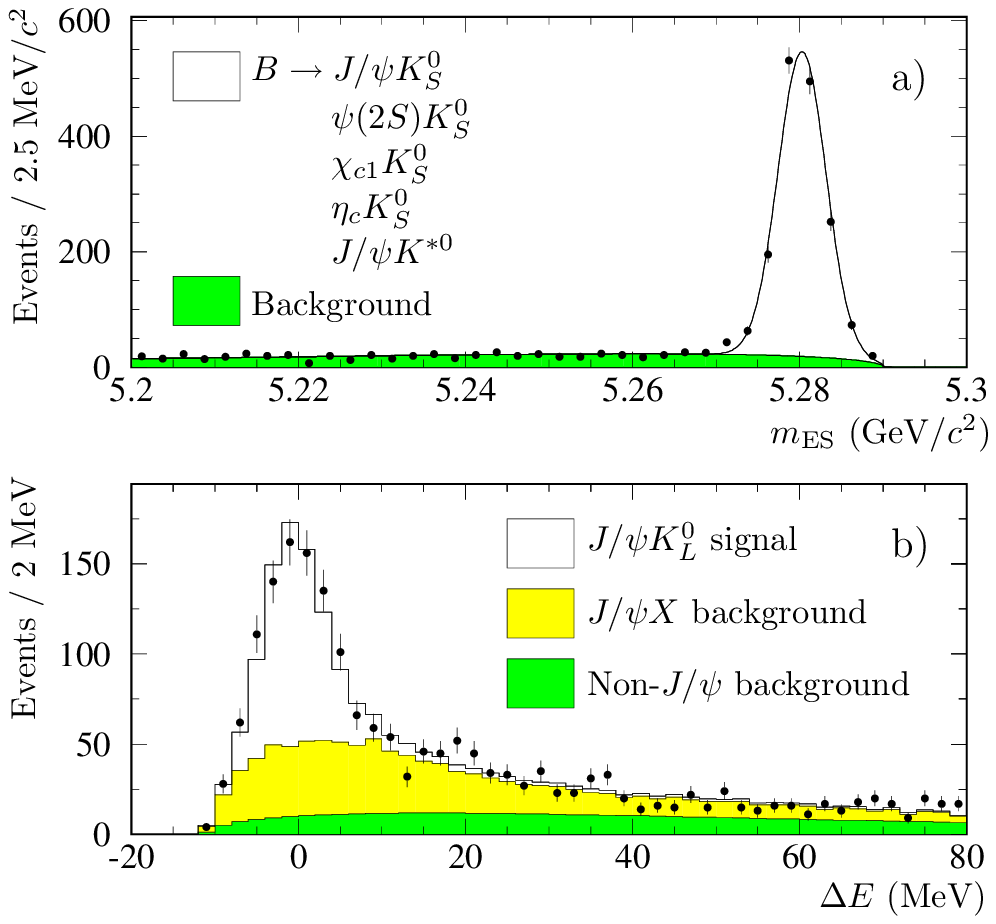}\quad
 \includegraphics[width=.48\linewidth]{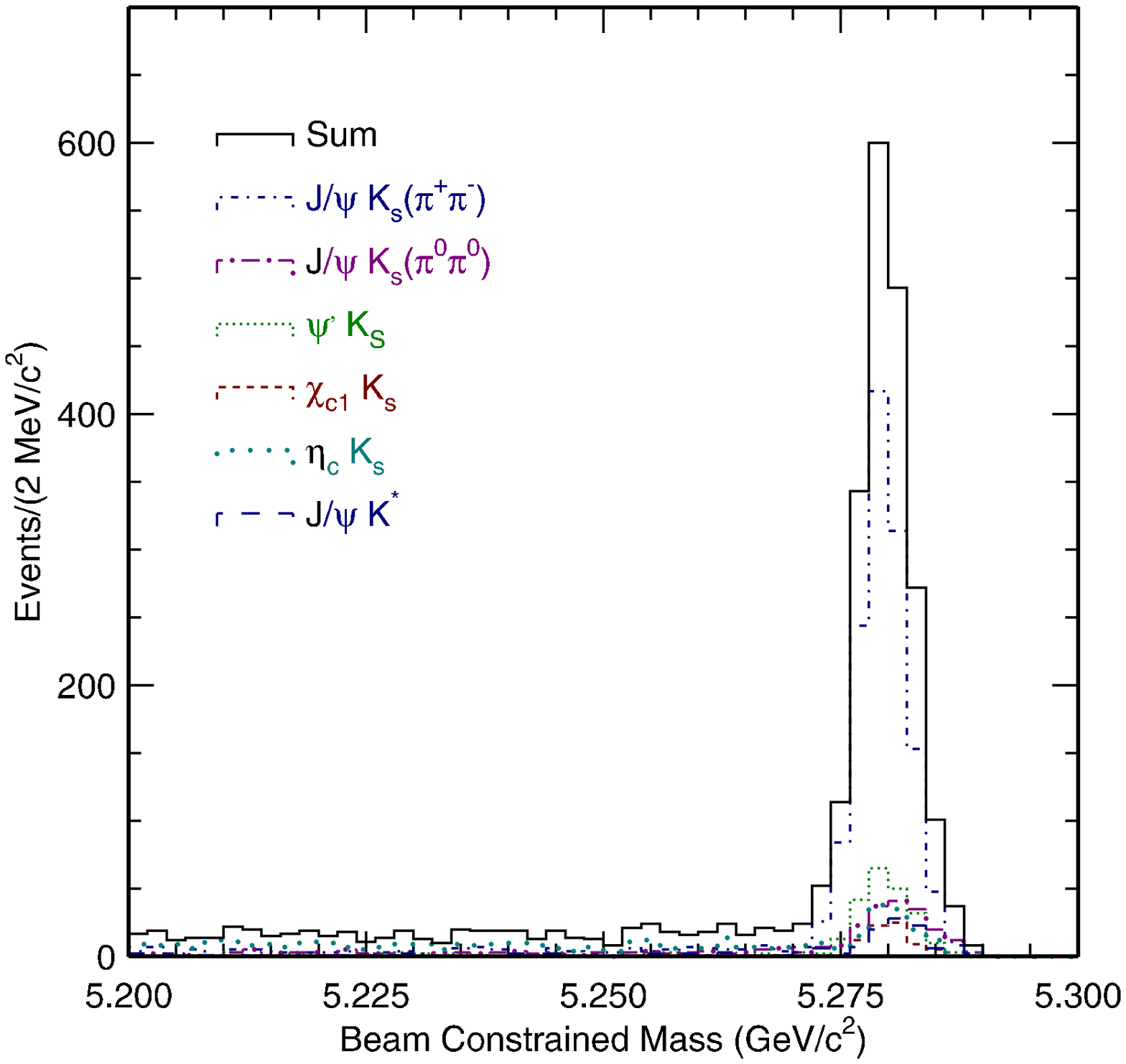}
 }
\caption{\label{fig:charmoniumKzMes}Distributions in kinematic
variables for selected events with flavor tags from \babar\ 
(left) and Belle (right).  For the \KS\ modes (\CP\ odd, right and
upper left) the \mes\ distribution is shown.  For \KL\ (\CP\ even,
lower left), the energy residual is given.
}
\end{figure}
\begin{figure}[htbp]
 \scalebox{1}[.9]{\includegraphics[width=.505\linewidth]{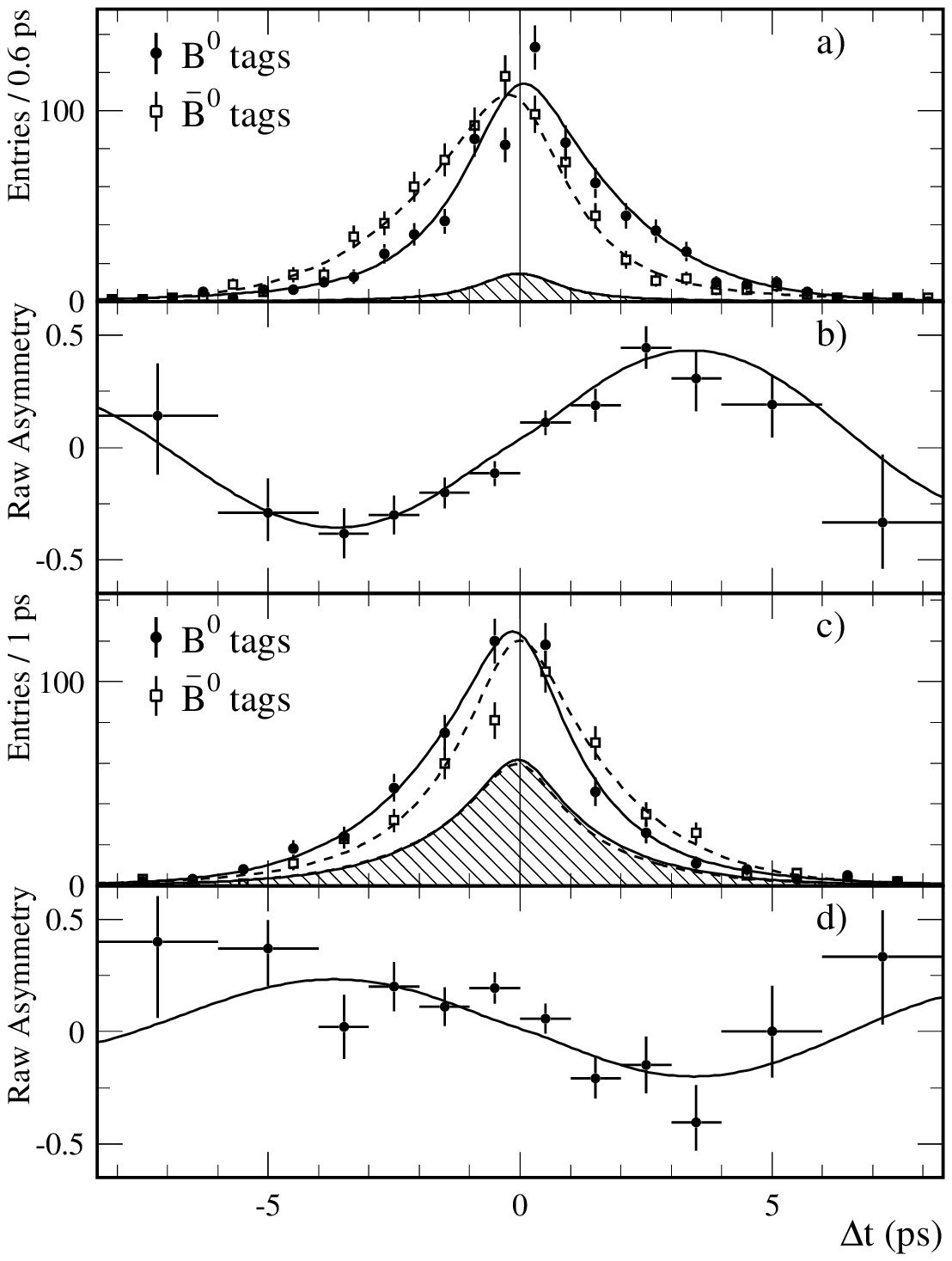}
  \quad\includegraphics[width=.47\linewidth]{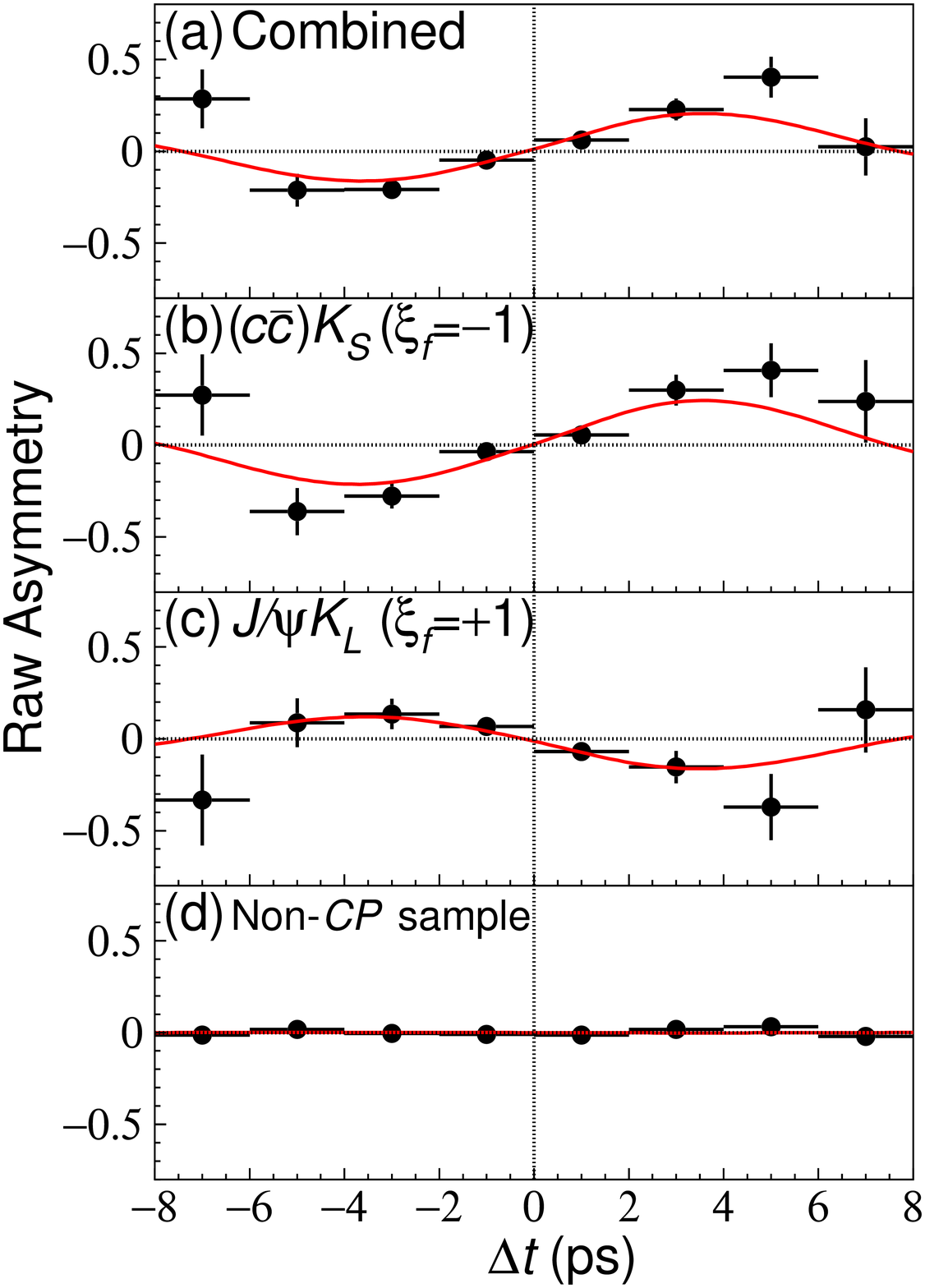}
}
\caption{\label{fig:charmoniumKzAsymResults}Rate and flavor asymmetry vs
\deltat\ from \babar\ (left) and Belle (right).  On the left appear
(a, c) the tagged-sample rates and (b, d) asymmetries for (a, b) \CP-odd and
(c, d) \CP-even modes.  On the right the asymmetries are given for (a) all
modes, for (b) \CP-odd ($\xi_f=\eta_f=-1$) and (c) \CP-even modes, and for
(d) a flavor-definite control sample.  
}
\end{figure}

\babar\ \cite{chmmK0BB}\ and Belle \cite{chmmK0Bl}\ have performed this
measurement with samples combining
several charmonium \KS\ modes, as well as $J/\psi\KL$ (for which
$\eta_f=+1$).  The contributing modes are listed in Fig.\
\ref{fig:charmoniumKzMes}\ which conveys a sense of the high purity of
these samples.  The exposures correspond to 88 (85) million produced
\BB\ pairs for \babar\ (Belle).
The \deltat\ distributions from both experiments are shown in Fig.\
\ref{fig:charmoniumKzAsymResults}.
From a 34-parameter likelihood fit to 2641 tagged events (with 78\%
purity), \babar\ measure \stwob\ given in Eq.\ \ref{eq:stwobBB}.  The
corresponding Belle measurement, from 2958 tagged events (having 81\%
purity), is given in Eq.\ \ref{eq:stwobBl}.
\beqa
  (\mathrm{with}\ \left|\lambdacp\right|=1)\ \stwob
         &=& 0.741\pm0.067\pm0.034\qquad{\rm\babar}\label{eq:stwobBB} \\
         &=& 0.719\pm0.074\pm0.035 \qquad{\rm Belle}
\eeqa{eq:stwobBl}
These results agree well with each other, and together impose a
significant new constraint on the upper vertex of the CKM triangle,
consistent with prior knowledge.

\section{\boldmath \stwob\ and probes for new physics with rare \B\ decays}\label{sec:rare}

Violations of \CP\ symmetry can be also observed in a number of rarer
\B\ decays.  In contrast with the ${\cal O}(\lambdackm^2)$ $b\ra
c\cbar s$ modes we've been discussing, these have couplings of ${\cal
O}(\lambdackm^3)$, or are expected to be suppressed because their
amplitudes contain penguin loops, or both.  We include among these the
decays with Cabibbo-suppressed $b\ra c\cbar d$ (e.g., charmonium \piz,
open charm pair) and the gluonic penguin $b\ra s\qbar q$ (e.g.,
$\phi\KS,\ \etapr\KS$).  These processes are sensitive to the presence
of possible new physics, because with their smaller amplitudes
interference terms are relatively more prominent, and because of
possible virtual particles (e.g., SUSY) in penguin loops.  These
experiments are harder because of lower rates, higher backgrounds, and
complications in their interpretation including the simultaneous
presence of tree and penguin amplitudes, multiple penguin amplitudes,
uncertainties from long-distance effects, etc.

In the time-evolution formalism for these decays we account for
effective direct and interference \CP-violating contributions by
removing the assumption $\alam=1$ and introducing
coefficients $S_f$ and $C_f$ of the sine- and cosine-like terms:
for a \CP\ eigenstate $f$ we have
\beq
     \frac{1}{\Gamma^f}\frac{d\Gamma^f_\pm(\deltat)}{d\deltat} =
\frac{e^{-\left|\deltat\right|/\tau}}{4\tau}  
     \left\{1 \mp\Delta w \pm (1-2\langle w \rangle)\left[
     {S_f}\sin(\deltamd\deltat) -
     {C_f}\cos(\deltamd\deltat)\right]\right\}\otimes{\cal R}, 
\eeq{eq:dtSC}
where
\beq
 S_f \equiv \frac{2\ilam}{1+\alam^2},\qquad
 -\calA_f = C_f \equiv \frac{1-\alam^2}{1+\alam^2},
\eeq{eq:SCdef}
$\Delta w = w(\Bz)-w(\Bzb)$, and as in section
\ref{sec:dt} we assume $\Gamma_H=\Gamma_L$.
For reference, $S_f=\stwob$ and $C_f=0$ (no direct \CP\ violation in
the decay) for $\Bz\ra J/\psi \KS$.

\subsection{\boldmath $b\goto c\cbar d$ decays}

\begin{figure}[htbp]
 \vspace{-5mm}
 \begin{center}
 \scalebox{.8}{
  (a) \includegraphics{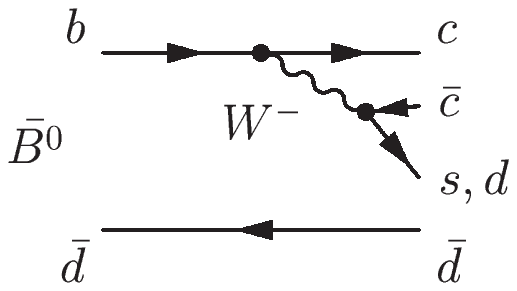}
  \qquad (b)\quad
   \scalebox{.6}{\includegraphics[width=.45\linewidth,bb=135 629 252 702,clip]{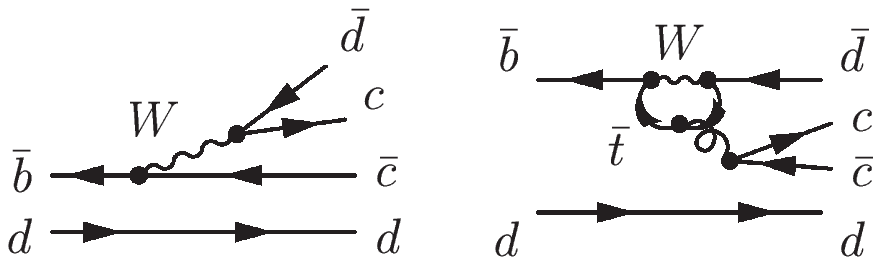}}
  \qquad (c)\quad
  \includegraphics{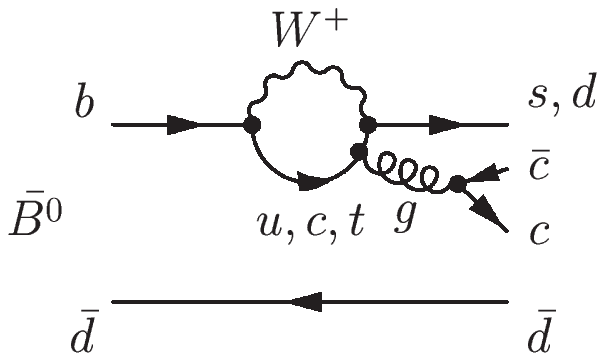}
 }
 \vspace{-8mm}
 \end{center}
 \caption{\label{fig:bccqDiagrams}Diagrams for $b\ra c \cbar q$
  decays; (a) color-suppressed tree; (b) external tree; (c) gluonic
 penguin.} 
\end{figure}

In section \ref{sec:chmmK}\ we considered only tree ($T$) decay
amplitudes.  In general we have a competition between amplitudes like
those shown in Fig.\ \ref{fig:bccqDiagrams}.  There are two
independent terms among the penguins ($P$) with $(u,c,t)$ in the loop,
and for the interpretation here we care about whether these bring in
weak phases different than that of the tree amplitude.  The
non-GIM-suppressed pieces are $P_c-P_t$, which has the same \CP\ phase
as $T$, and $P_u-P_t$, which has a different \CP\ phase.  For
$b\ra c\cbar s$, such as $J/\psi\KS$, the ratio of $P_u-P_t$ to $T$ is
${\cal O}(\lambdackm^4/\lambdackm^2)$, justifying our neglect of it.
But for Cabibbo-suppressed $b\ra c\cbar d$ decays such as
$J/\psi\pi^0$, both are ${\cal O}(\lambdackm^3)$, and thus the $P/T$
ratio becomes a theoretical systematic for the interpretation of
$b\ra c\cbar d$ decays.

With this caveat, we quote the results of measurements
in the $b\ra c\cbar d$ counterpart $\Bz\goto J/\psi\pi^0$ of the
charmonium-\KS\ decays:
\babar\ \cite{chmmPi0BB}, with $40\pm7$ signal events from $88\times10^6\ \BB$
pairs find
   \beq
    S_{J/\psi\pi^0} = 0.05\pm0.49\pm0.16 \qquad
    C_{J/\psi\pi^0} = 0.38\pm0.41\pm0.09,
   \eeq{eq:psiPi0bb}
and Belle \cite{chmmPi0Bl}, with 57 total events (86\% purity) from
$85\times10^6\ \BB$ measure 
   \beq
    S_{J/\psi\pi^0} = 0.93\pm0.49\pm0.11^{+0.27}_{-0.03} \qquad
    -C_{J/\psi\pi^0} = {\cal A} = -0.25\pm0.39\pm0.06.
   \eeq{eq:psiPi0bl}
Both measurements are preliminary.  They are consistent with
$S_{J/\psi\pi^0}=\stwob, C_{J/\psi\pi^0}=0$, within large errors.


\babar\ have measured the time evolution for two modes with open charm
pairs.  Again the quark-level process is $b\ra c\cbar d$ (and so 
$P_u-P_t$ again brings in a second weak phase), although there is in
this case no color-suppression of the tree (see Fig.\
\ref{fig:bccqDiagrams}b); the estimate is $\Delta\beta\sim0.1$, where
$\Delta\beta$ is the deviation of the measured $\beta_\mathrm{eff}$
from true $\beta$.

%
%
The $D^{*\pm}D^\mp$ decays
are not \CP\ eigenstates, but are
accessible from \Bz\ and \Bzb.  The decay chains analyzed are
$D^{*\pm}\ra\pi^\pm D^0$ with four $D^0$ modes, and $D^+\ra
K\pi\pi,\KS\pi$.  With $113\pm13$ signal events from $88\times10^6\
\BB$ pairs the results are \cite{dstarDBB}
\beqa
    S_{-+} = -0.24\pm0.69\pm0.12 &\quad&
    C_{-+} = -0.22\pm0.37\pm0.10 \\
    S_{+-} = -0.82\pm0.75\pm0.14 &\quad&
    C_{+-} = -0.47\pm0.40\pm0.12.
\eeqa{eq:dstarD}
Here $S_{+-}$ corresponds to $D^{*+}D^-$, etc., and
if one assumes equal amplitudes for $D^{*-}D^+$ and $D^{*+}D^-$ one expects
$C_{-+}=C_{+-}=0$; if penguins are negligible $S_{-+}=S_{+-}= -\stwob=-0.7$.

The same quark-level amplitudes describe $\Bz\goto
D^{*\pm}D^{*\mp}$.  This, however, is a vector-vector decay, with
odd-\CP\ $P$-wave and even-\CP\ $S$- and $D$-wave contributions.  
%
%
The final state is reconstructed in the chains $D^{*\pm}\ra
D^0\pi^\pm,D^\pm\pi^0$, excluding $B^0\ra D^+D^-\pi^0\pi^0$.  An
angular analysis \cite{dstarDstarBB}
yields for the \CP-odd fraction $R_\perp=0.063\pm0.055\pm0.009$, i.e., 
this final state is $\sim94\%$ \CP-even.  
With $156\pm14$ signal events (before tagging, 
73\%\ purity) from $88\times10^6\ \BB$ pairs, \babar\ find the
preliminary results
\beq
    \left|\lambdacp_+\right| = 0.75\pm0.19\pm0.02, \qquad
    {\cal I}m\lambdacp_+ = 0.05\pm0.29\pm0.10,
\eeq{eq:CPdstarDstar}
where $\lambdacp_+$ refers to the \CP-even component.  We may compare
this with the tree-level expectation $\left|\lambdacp_+\right|=1$ and
${\cal I}m\lambdacp_+ = -\stwob$.

\subsection{\boldmath $b\goto s\qbar q$ decays}
\begin{figure}[htbp]
  \scalebox{1}[.71]{
   \includegraphics[width=0.48\linewidth]{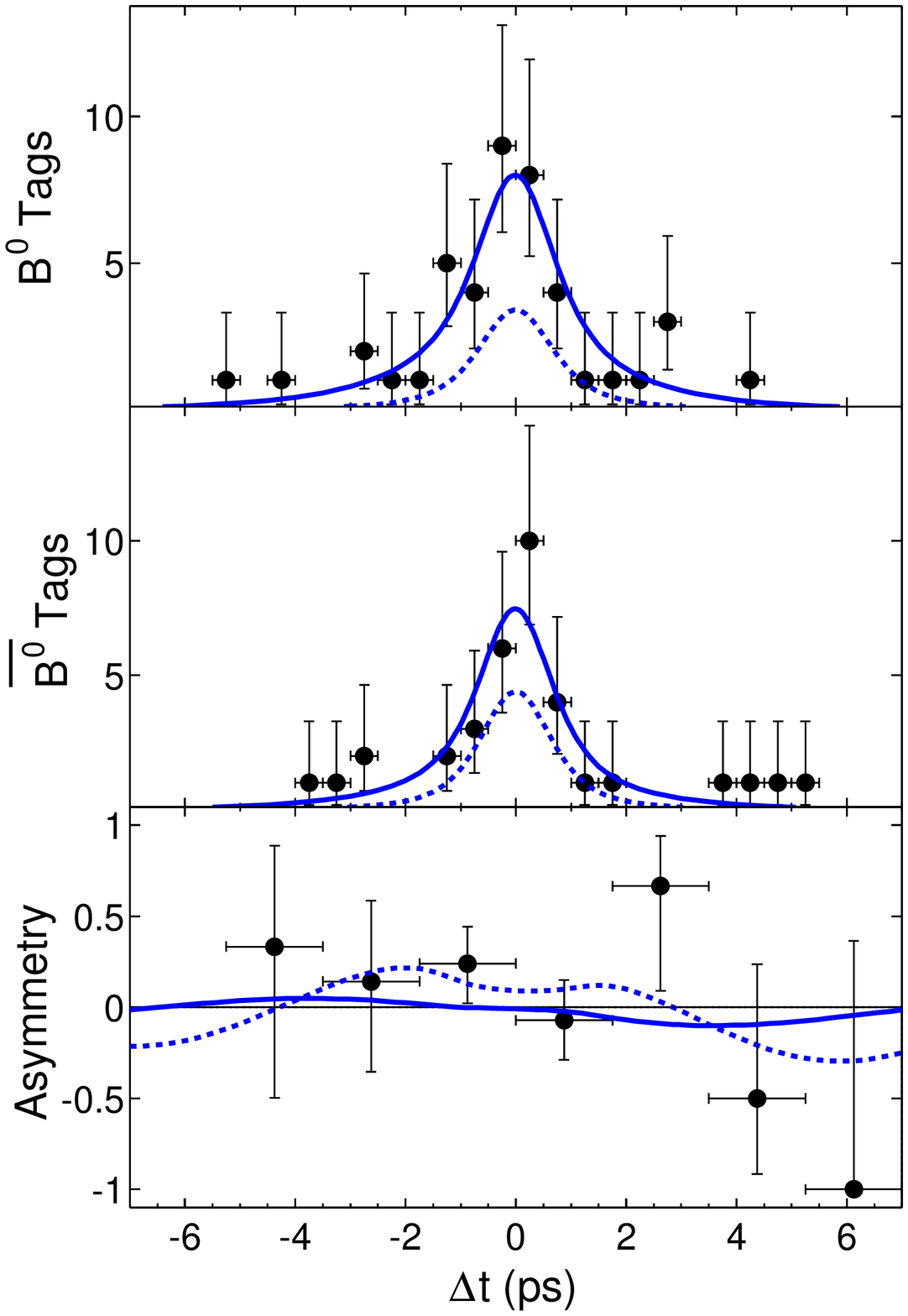}
  }\quad
  \scalebox{1}[1]{\includegraphics[width=.525\linewidth]{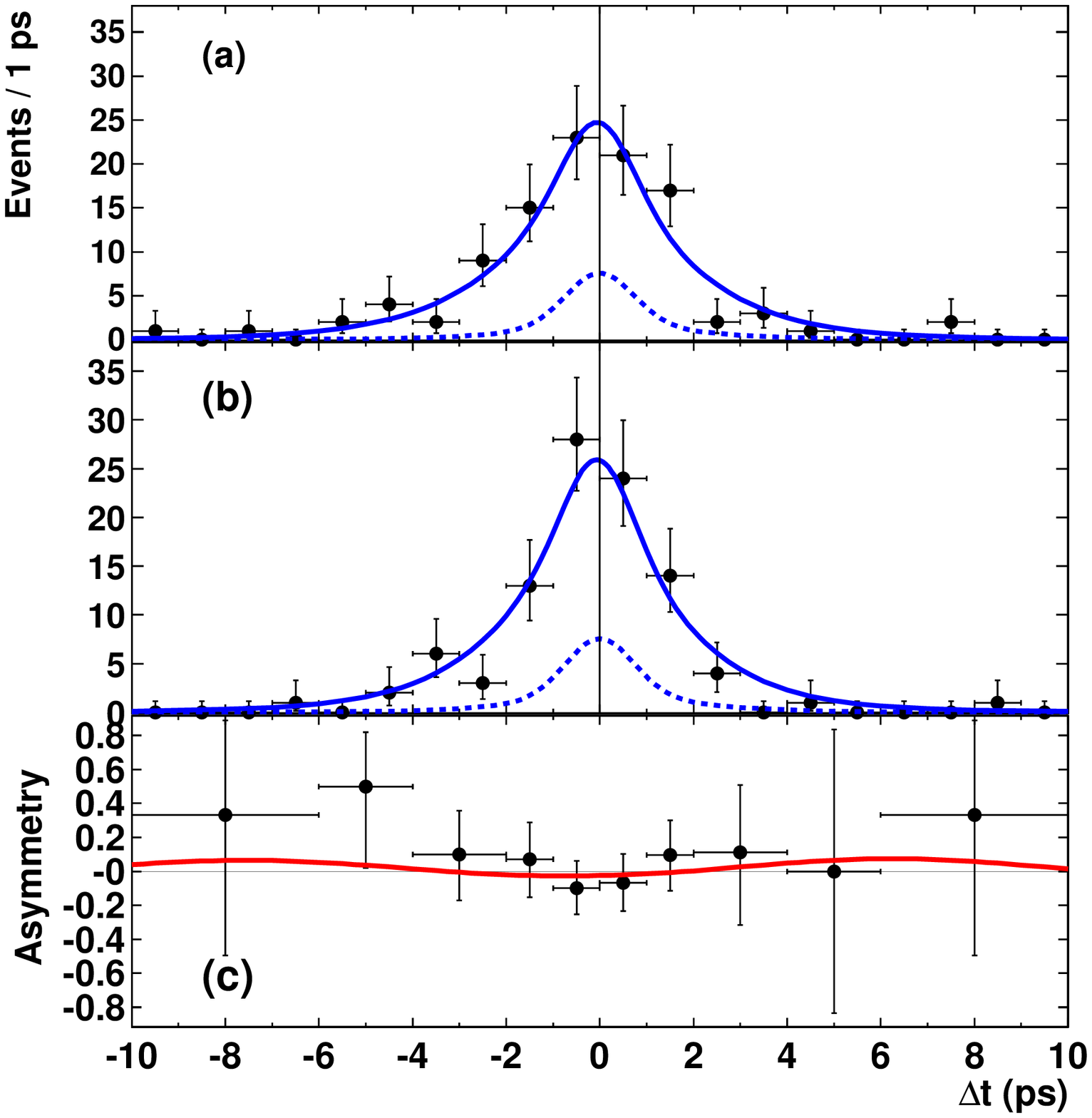}}
 \caption{\label{fig:phiK0bb}Tagged-sample \deltat\ distributions and
  asymmetry for (left) $\Bz\ra\phi\KS$ and 
  (right) $\Bz\ra\etapr\KS$ (\babar, preliminary).  Tags in the
right-hand plot are (a) \Bz\ and (b) \Bzb.  Dashed curves show
   background components of the best-fit functions.}
\end{figure}

For $b\goto s\qbar q$ decays with no $c$-quarks in the final state, the
tree is a CKM-suppressed $b\ra u$ with both color and Cabibbo
suppression at the internal vertex.  Therefore the leading amplitudes
are gluonic $b\ra s$ penguins.  The ratio $T/P$
is ${\cal O}(\lambdackm^4/\lambdackm^2)$.  For $\Bz\ra\etapr\KS$ the
internal gluon converts to either an $s\sbar$ or $d\dbar$ pair, and
these may interfere.  The rather large ($60\times10^{-6}$) branching
fraction for this decay may in fact be the result of constructive
interference of these amplitudes.  In the case of $\phi\KS$ we
have no tree, and only the penguins with $g\ra s\sbar$.  Estimates of
$\Delta\beta$ from ``tree polution'' are as small as 0.01 (0.025) for
$\Bz\ra\etapr\KS$ ($\phi\KS$) \cite{dBeta_sqqThy}. 

\begin{figure}[htbp]
 \scalebox{1}{
  \includegraphics[width=.31\linewidth]{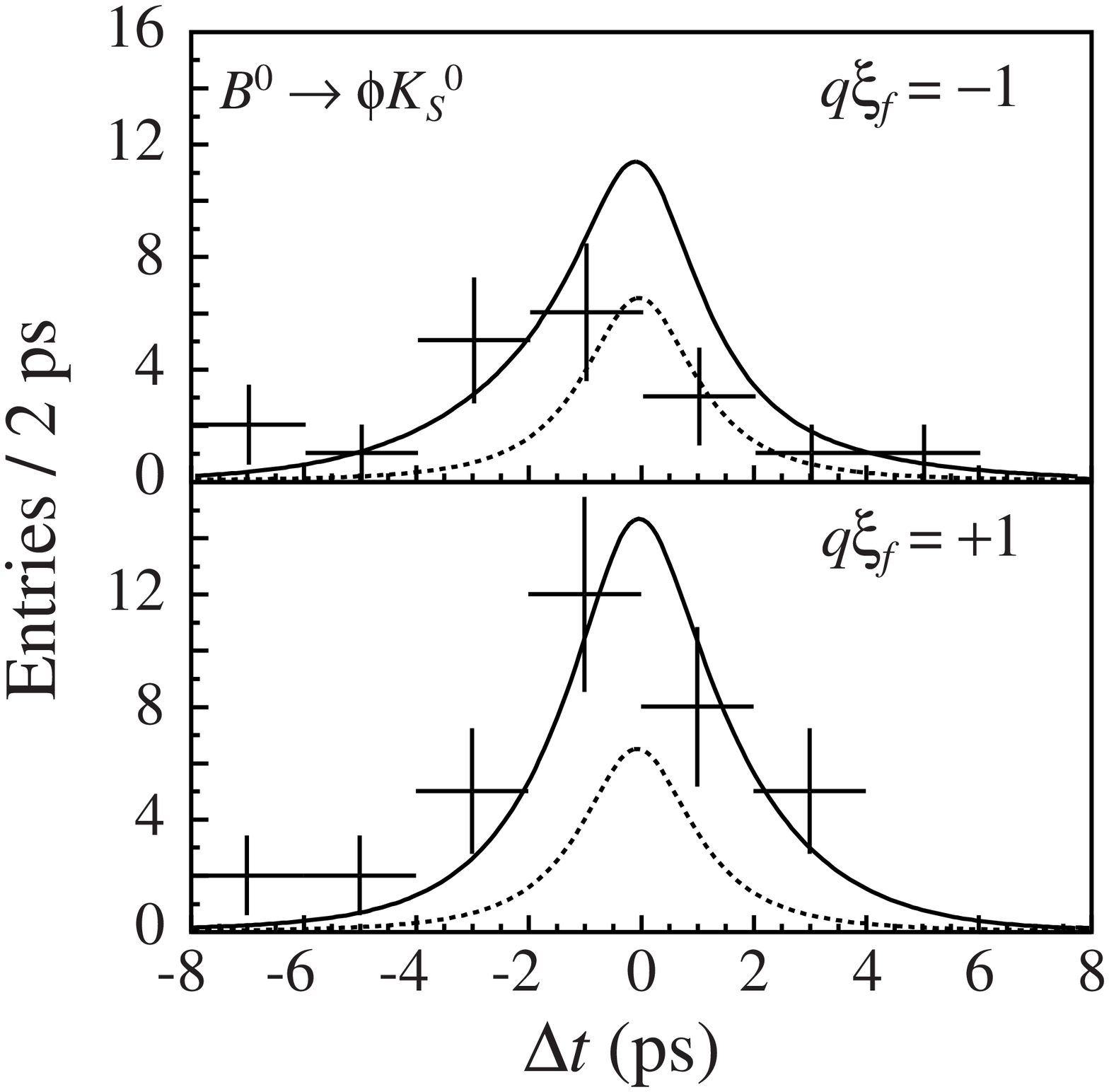}
  \quad\includegraphics[width=.31\linewidth]{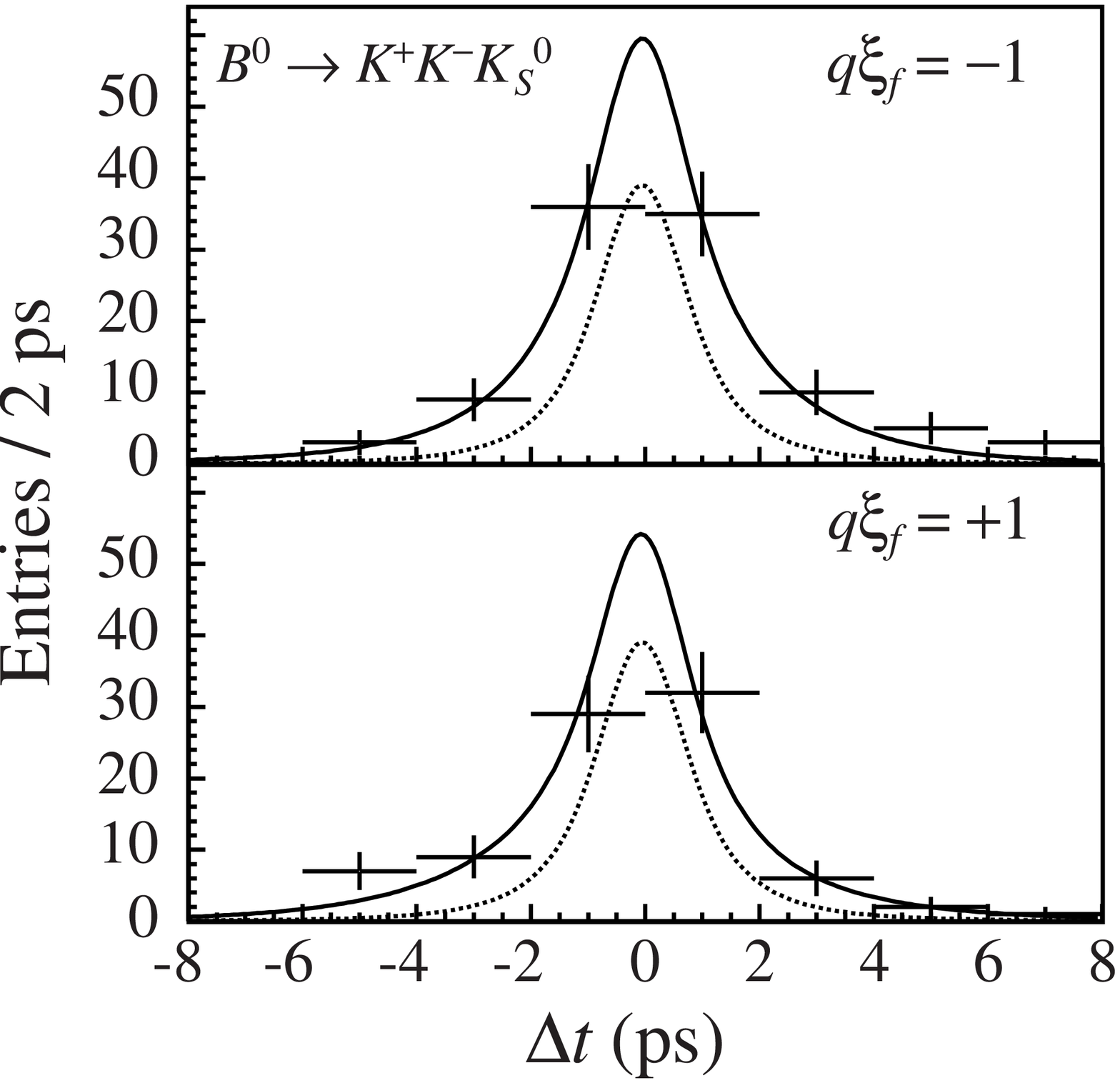}
  \quad\includegraphics[width=.31\linewidth]{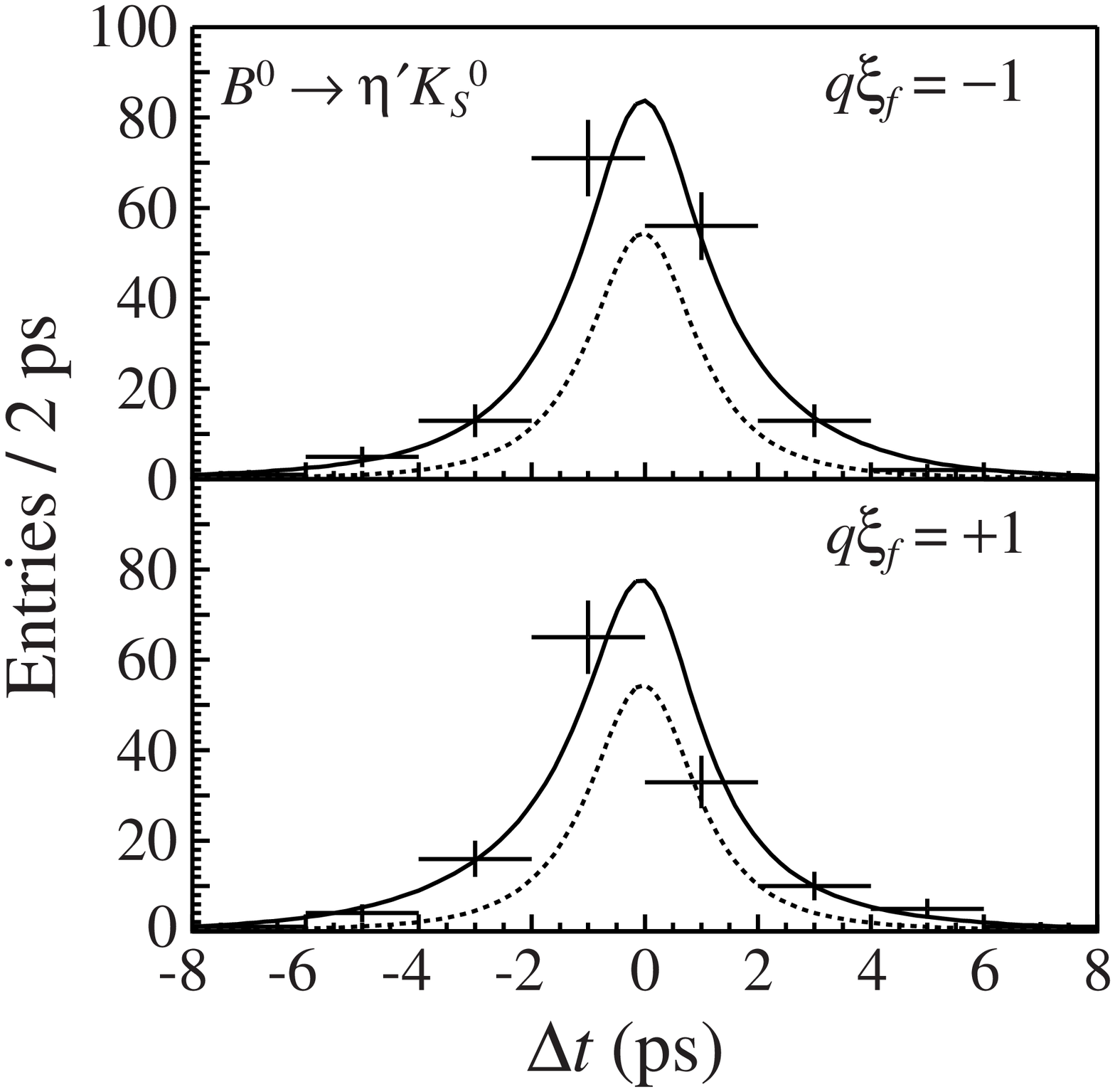}
 }
 \caption{\label{fig:sqqBelle}Tagged \Bz\ ($q=+1$) and \Bzb\ ($q=-1$)
  \deltat\ distributions for $b\ra s q\qbar$ decays (Belle).}
\end{figure}

%
For $\Bz\ra\phi\KS$ the $\phi$ is reconstructed from its \Kp\Km\ decay,
and the \KS\ from 
\pip\pim\ (\babar\ include also \piz\piz).  With 51 signal events (31
tagged), from $87\times10^6\ \BB$ pairs the \babar\ preliminary results
are \cite{phiKsBB}
\beq
    S_{\phi\KS} = -0.18\pm0.51\pm0.07\qquad
    C_{\phi\KS} = -0.80\pm0.38\pm0.12,
\eeq{eq:phiKsBB}
($S_{\phi\KS}=-0.26\pm0.51$ when $C_{\phi\KS}$ is constrained to zero).
Belle find with 53 total events (purity 67\%) from $78\times10^6\ \BB$
\cite{bsqqBl}
\beq
  S_{\phi\KS} = -0.73\pm0.64\pm0.22\qquad
  {\cal A}_{\phi\KS} = -C_{\phi\KS} = -0.56\pm0.41\pm0.16.
\eeq{eq:phiKsBl}
The \deltat\ distributions are shown in the left-hand plots in Figs.\
\ref{fig:phiK0bb} and \ref{fig:sqqBelle}, respectively.
Taken together these are somewhat at odds with the expectation
$S_{\phi\KS}=\stwob=0.7$; the errors are still large however.

From Belle there is also a measurement in non-resonant $\Bz\goto
\Kp\Km\KS$, a mixture of \CP-odd and even states.  Their analysis
indicates that it is in fact about 97\%\ even ($\xi_f=+1$), and with their
191 total events (purity 50\%) they find (see center plot
of Fig.\ \ref{fig:sqqBelle}) \cite{bsqqBl} 
\beq
     -\xi_fS = 0.49\pm0.43\pm0.11^{+0.33}_{-0.00}\qquad
     {\cal A} = -C = -0.40\pm0.33\pm0.10^{+0.00}_{-0.26}.
\eeq{eq:KKKsBl}

The $\Bz\goto\etapr \KS$ decay is reconstructed from
$\etapr\ra\eta\pip\pim$ and $\etapr\ra\rho^0\gamma$.
With Belle's sample of 299 total events (purity 49\%) the measurements
are (right-hand plot of Fig.\ \ref{fig:sqqBelle}) \cite{bsqqBl}
\beq
     \skz = +0.71\pm0.37^{+0.05}_{-0.06} \qquad
     {\cal A} = -\ckz = +0.26\pm0.22\pm0.03
\eeq{eq:etaprKsBl}
(consistent with the charmonium value of \stwob), and
%
from \babar, with 109 tagged signal events (purity 70\%) from
$89\times10^6\ \BB$ pairs (right-hand plot of Fig.\ \ref{fig:phiK0bb})
\cite{etapKsBB} 
\beq
     \skz = \rSetapKz, \quad \ckz = \rCetapKz.
\eeq{eq:etaKsBB}
This (preliminary) null result is still consistent within errors with
$\stwob=0.7$. 

\begin{figure}[htbp]
 \begin{center}
  \includegraphics[width=.6\linewidth]{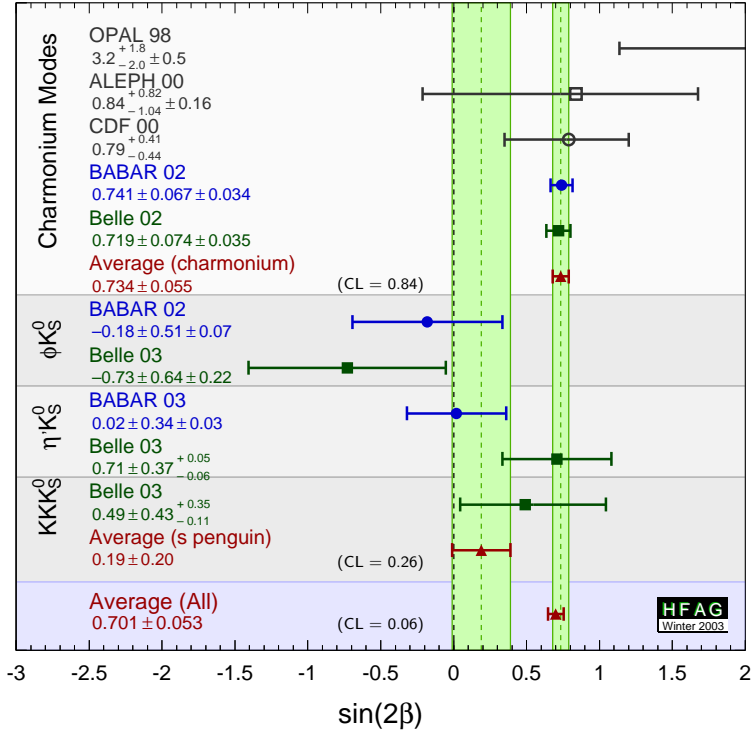}\\[-64.5mm]
\tiny{{\sf (CL = 0.84)}}\\[34.5mm]
\tiny{{\sf (CL = 0.26)}}\\[5.8mm]
\tiny{{\sf (CL = 0.06)}}\\[12mm]
 \end{center}
 \caption{\label{fig:wasin2b_c}Measurements of \stwob\ and their world
averages.} 
\end{figure}

\section{Summary}

All of the measurements presented here are summarized in Fig.\
\ref{fig:wasin2b_c}\ \cite{HFAG}.  The shaded bands show results of
averaging separately the measurements made with \B\ decays to charmonium
and with the charmless penguin-dominated modes.  Future confirmation
of the separation of the two bands would challenge the standard model,
but considerable improvement in the theoretical understanding as well
as in the experimental measurements would be needed for a
definitive conclusion.

We have seen that \CP\ non-conservation is well established in \Bz\
decays, and the effect is large in the interference between mixing and
decay.  The effect is well accommodated in the universal weak
interactions of the quarks embodied in the standard CKM model.  We are
just beginning to explore further whether anything new is happening in
processes where standard model effects are suppressed.  We see hints of
inconsistencies in $\beta$ from charmless decays, where several channels
are now being measured, with only more data needed for definitive
results.

\section*{Acknowledgments}
It is a pleasure to thank the organizers of the conference for providing
an enlightening program in a most hospitable setting.  This work was
supported in part by the Department of Energy under grant
DE-FG03-95ER40894.

%
\label{FordEnd}
 
\end{document}